\documentclass{iopjournal}

\usepackage{caption}
\usepackage{float}
\usepackage{url}
\usepackage{xcolor}
\usepackage{etoolbox}
\usepackage[style=numeric, sorting=none]{biblatex}
\begin{document}

%\articletype{Paper} %	 e.g. Paper, Letter, Topical Review...

\title{Gyrokinetic Simulations of a Low Recycling Scrape-off Layer without a Lithium Target}

\author{A. Shukla$^{1,2,*}$\orcid{0000-0002-2565-548X},
    J. Roeltgen$^{1,2}$\orcid{0000-0003-2418-0221},
    M. Kotschenreuther$^{2}$\orcid{0000-0001-6327-877X},
    D. R. Hatch$^{1,2}$\orcid{0000-0002-1625-4385},
    M. Francisquez$^3$\orcid{0000-0002-8247-3770},
    J. Juno$^3$\orcid{0000-0001-6835-273X},
    T. N. Bernard$^4$\orcid{0000-0002-7331-9704},
    A. Hakim$^3$\orcid{0000-0001-6603-8595},
    G. W. Hammett$^3$\orcid{0000-0003-1495-6647},
    and S. M. Mahajan$^{1,2}$\orcid{0000-0003-2415-4840}}

\affil{$^1$Institute for Fusion Studies, The University of Texas at Austin, Austin, USA}

\affil{$^2$Exofusion, Bellevue, USA}

\affil{$^3$Princeton Plasma Physics Laboratory, Princeton, USA}

\affil{$^4$General Atomics, San Diego, USA}

\affil{$^*$Author to whom any correspondence should be addressed.}

\email{akashukla@utexas.edu}

\keywords{Fusion plasma, Low recycling, Divertor, Tokamak}

\begin{abstract}
Low-recycling regimes are appealing because they entail a high edge temperature and low edge density which are good for core confinement. 
However, due to considerably enhanced heat flux, the exhaust problems become severe. In addition, in the low-recycling regime, the conventional fluid simulations may not capture the physics of the Scrape-Off Layer (SOL) plasma that lies in the long mean free path regime; kinetic calculations become necessary. In this paper, by performing both Kinetic and fluid simulations, we explore the feasibility of a low-recycling regime in the magnetic geometry of the Spherical Tokamak for Energy Production (STEP); kinetic effects come out to be crucial  determinants of the SOL dynamics.
The simulation results indicate that a high SOL temperature and low SOL density could be achieved
even when the divertor target is not made of a low recycling material. This can be done by using a low recycling material as a wall material.
This is an important step towards demonstrating the feasibility of a low-recycling scenario.
Lithium, a commonly used low recycling material, tends to evaporate at high heat fluxes which counteracts the desired high temperature, low density regime, and materials that can handle high heat fluxes are generally high recycling. 
Comparisons of gyrokinetic and fluid simulation results indicate that one can take advantage of kinetic effects to address some of the issues associated with a low-recycling SOL. Specifically, kinetic simulations show better confinement of impurities to the divertor region and greater broadening of the heat flux width due to drifts when compared with fluid simulations. Impurity confinement would help prevent core contamination from sputtering, and a broader heat flux width would help reduce the peak heat load at the target in the absence of detachment.
\end{abstract}

\section{Introduction}
Low-recycling regimes are appealing because they entail a high edge temperature and low edge density which are good for core confinement~\cite{Mike23}. 
However, present low recycling scenarios present challenges for heat flux handling and avoiding undue evaporation of lithium. Lithium tends to evaporate quickly at high heat fluxes and materials which handle high heat fluxes with less evaporation are high recycling. 
In addition to physical challenges, there are also modeling challenges associated with the low-recycling regime. Fluid simulations are typically used to study the Scrape-Off Layer (SOL) in tokamaks ~\cite{Hudoba2023, Osawa2023, Rozhansky2021, Zhang2024},  but modeling the collisionless SOL of a low recycling regime in which the fluid assumptions are not valid requires a kinetic treatment. 
Here we explore the feasibility of a low-recycling SOL scenario in the magnetic geometry of STEP. We also compare gyrokinetic and fluid simulations to investigate kinetic effects that can be taken advantage of to address the challenges of sputtering and heat flux handling in the low-recycling regime. We use Gkeyll~\cite{Shukla25,Shukla25Xpt,Mana25,antoine25} for gyrokinetic simulations and SOLPS~\cite{Wiesen25,Schneider2006} for fluid simulations. The simulation results presented here indicate that (1) A high SOL temperature can be achieved without using a low-recycling material at the target but using it at the side-walls instead, (2) At high SOL temperatures, impurities can be electrostatically confined to the divertor region, and (3) The heat flux width is significantly broadened by mirror trapping.   

Lithium is a commonly used low recycling material but its high vapor pressure makes it unable to go above $\sim$ 400 $^\circ$C without evaporating, which tends to counteract the desired physical regime. Keeping the target temperature below 400 $^\circ$C is difficult, so it would be difficult to maintain a low temperature, high density SOL using lithium as a target material. The simulations presented in section~\ref{sec:10kev} indicate that a low-recycling SOL can be achieved by coating the side walls of the tokamak with Lithium but using a material that can handle high heat fluxes without evaporating, such as refractory solid metals or novel liquid metals~\cite{ARPAE2025, INFUSE2024, SuperXT}, on the target. This addresses a major concern about the feasibility of a low-recycling SOL.

In a low-recycling regime, the high heat fluxes at the target will entail a large amount of sputtering. One concern is that the sputtered impurities will contaminate the core plasma. The kinetic simulations shown in section~\ref{sec:impurities} indicate that, in less collisional regimes, impurities are much better confined to the divertor region than is predicted by fluid simulations. If the SOL temperature is high, the potential drop along the divertor leg can be very large which serves as a barrier preventing impurities from traveling upstream.

Fluid simulations used to model collisional SOLs typically neglect the mirror force. Our gyrokinetic simulation results in section~\ref{sec:drifts} show that in collisionless regimes, the effect of particle drifts in combination with mirror trapping has a large effect on the heat flux width. When drifts are included, kinetic simulations show that the heat flux width is broadened to approximately the ion banana width, while the broadening observed in fluid simulations is negligible. In a low-recycling SOL, the banana width can be quite large which can result in significant increase in the heat flux width and corresponding reduction in the peak heat flux.

\section{Gkeyll-EIRENE Simulations of a Low Recycling Scenario}
\label{sec:10kev}
\begin{figure}[H]
\begin{center}
\includegraphics[width=0.4\textwidth]{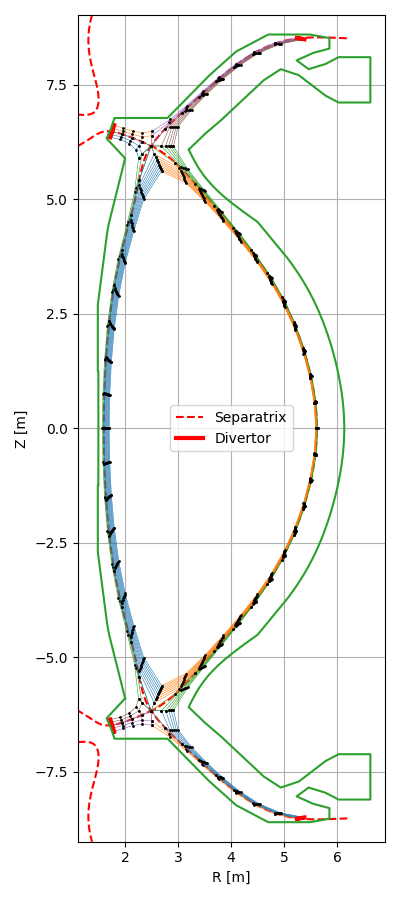}
\end{center}
\vspace{-0.5cm}
\caption{Simulation domain used for the coupled Gkeyll-EIRENE simulation. The colored grid lines show the Gkeyll simulation domain and the thick green boundary enclosing the Gkeyll domain indicates the machine boundary which is the EIRENE simulation domain.
\label{fig:domain} }
\end{figure}

We conducted a simulation of a low-recycling SOL using Gkeyll's axisymmetric gyrokinetic solver~\cite{Shukla25} coupled to the monte-carlo neutral code EIRENE~\cite{Wiesen25}. Gkeyll evolves the plasma and EIRENE evolves the neutral particles. These simulations are similar in concept to SOLPS simulations which couple the fluid solver B2.5 to EIRENE, but B2.5 has been replaced with a gyrokinetic solver to include kinetic effects. The simulation grid is shown in Fig.~\ref{fig:domain}. The simulation grid used in Gkeyll is shown, and the green boundary indicates the machine wall, which is the domain used for EIRENE.
\setlength{\parindent}{1.5em}

The simulation is sourced by Maxwellian with a temperature of 10keV at the inner radial boundary with an input power of 100 MW. The perpendicular particle and heat diffusivities are 0.22 $m^2/s$ and 0.33 $m^2/s$ respectively, and drifts are turned off. The diffusivities were chosen to target a heat flux width (mapped upstream) of 2 mm. The divertor plate has a recycling coefficient of 0.99 and the side walls are coated with lithium giving them a recycling coefficient of 0.5. The simulation evolves 3 species: deuterium ions, electrons, and neutral deuterium. The input files containing details of the simulation setup can be found at \url{https://github.com/ammarhakim/gkyl-paper-inp/tree/master/2025_IAEA_STEP}.

\begin{figure}[htbp]
\centering
\includegraphics[width=0.24\textwidth]{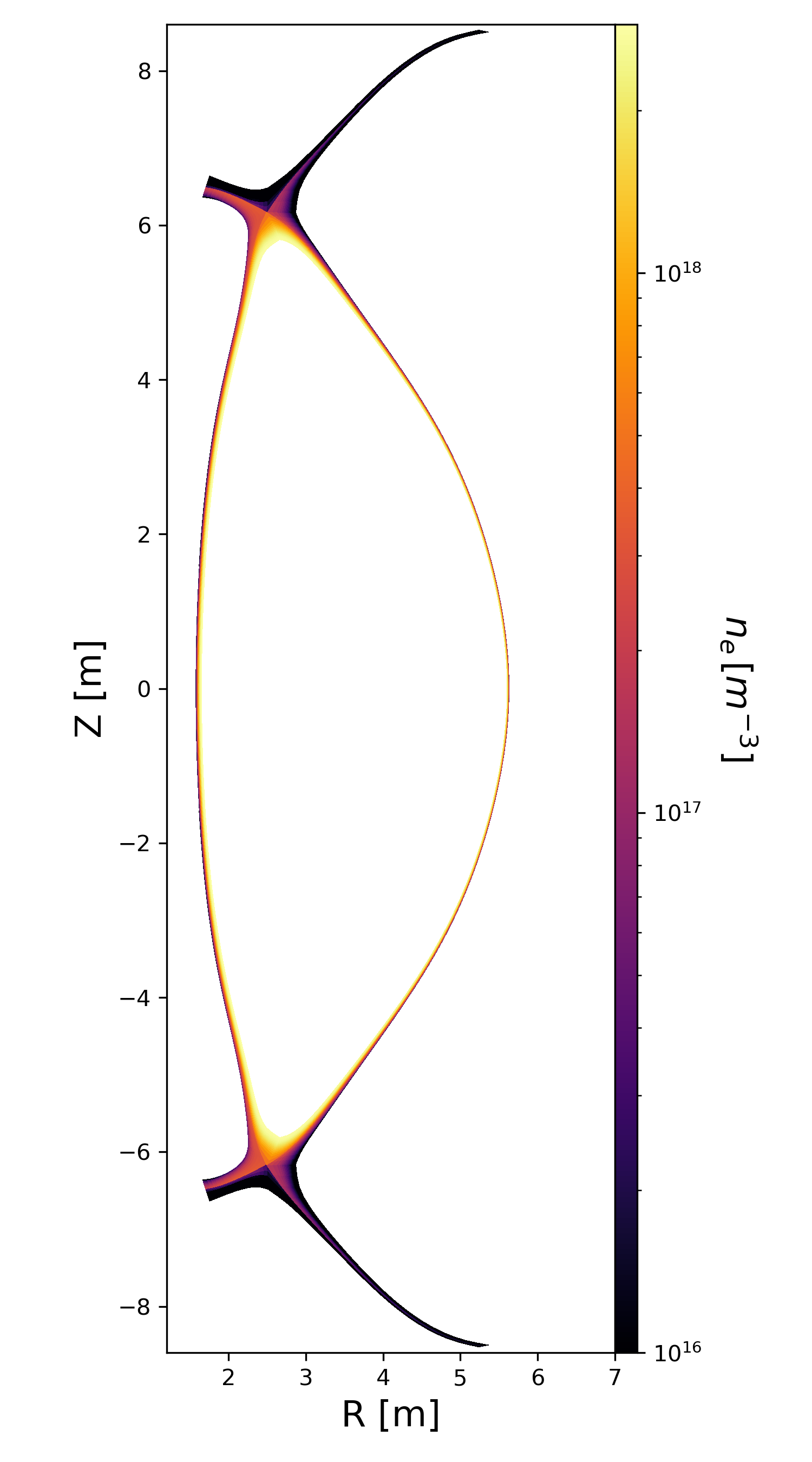}
\includegraphics[width=0.24\textwidth]{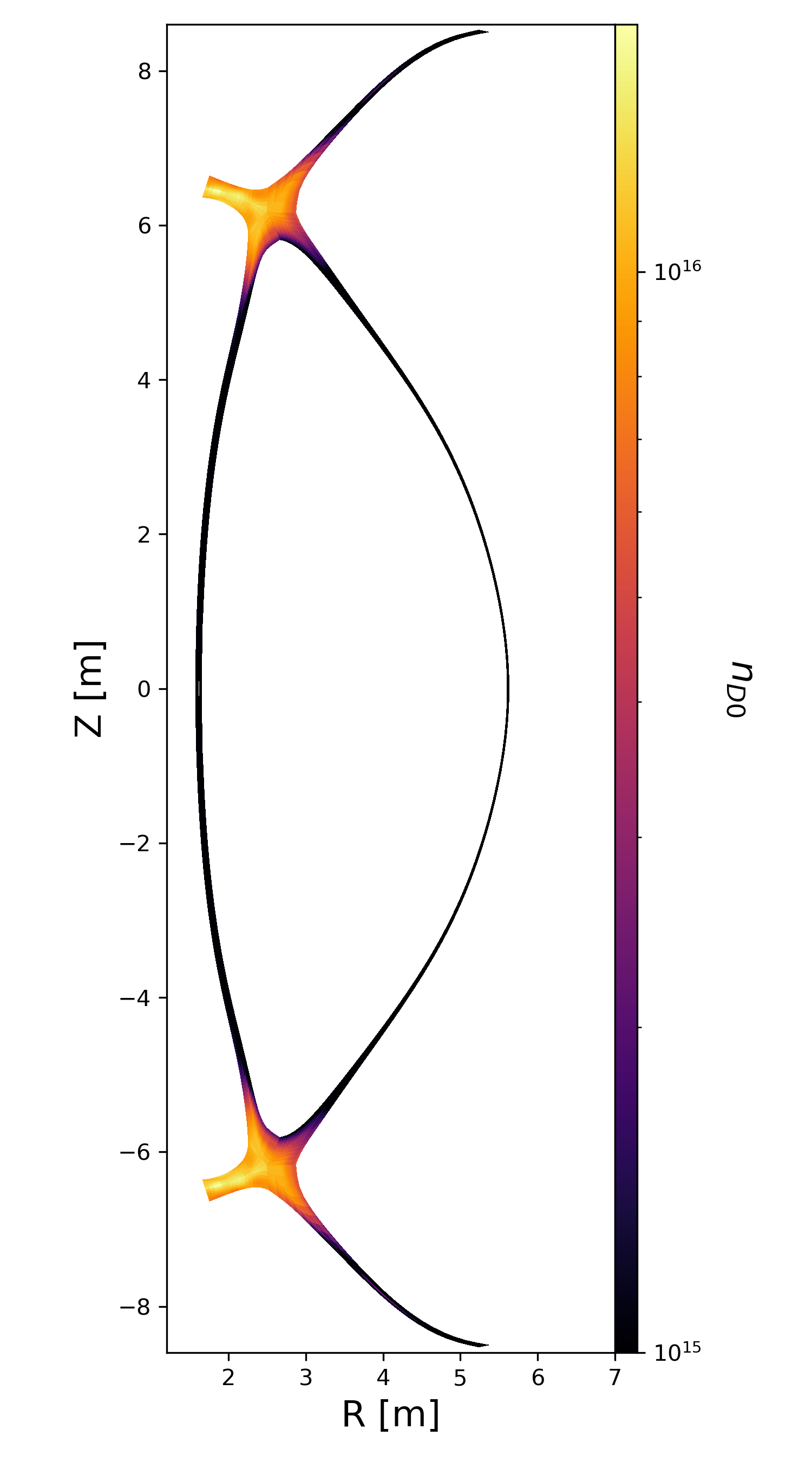}
\includegraphics[width=0.24\textwidth]{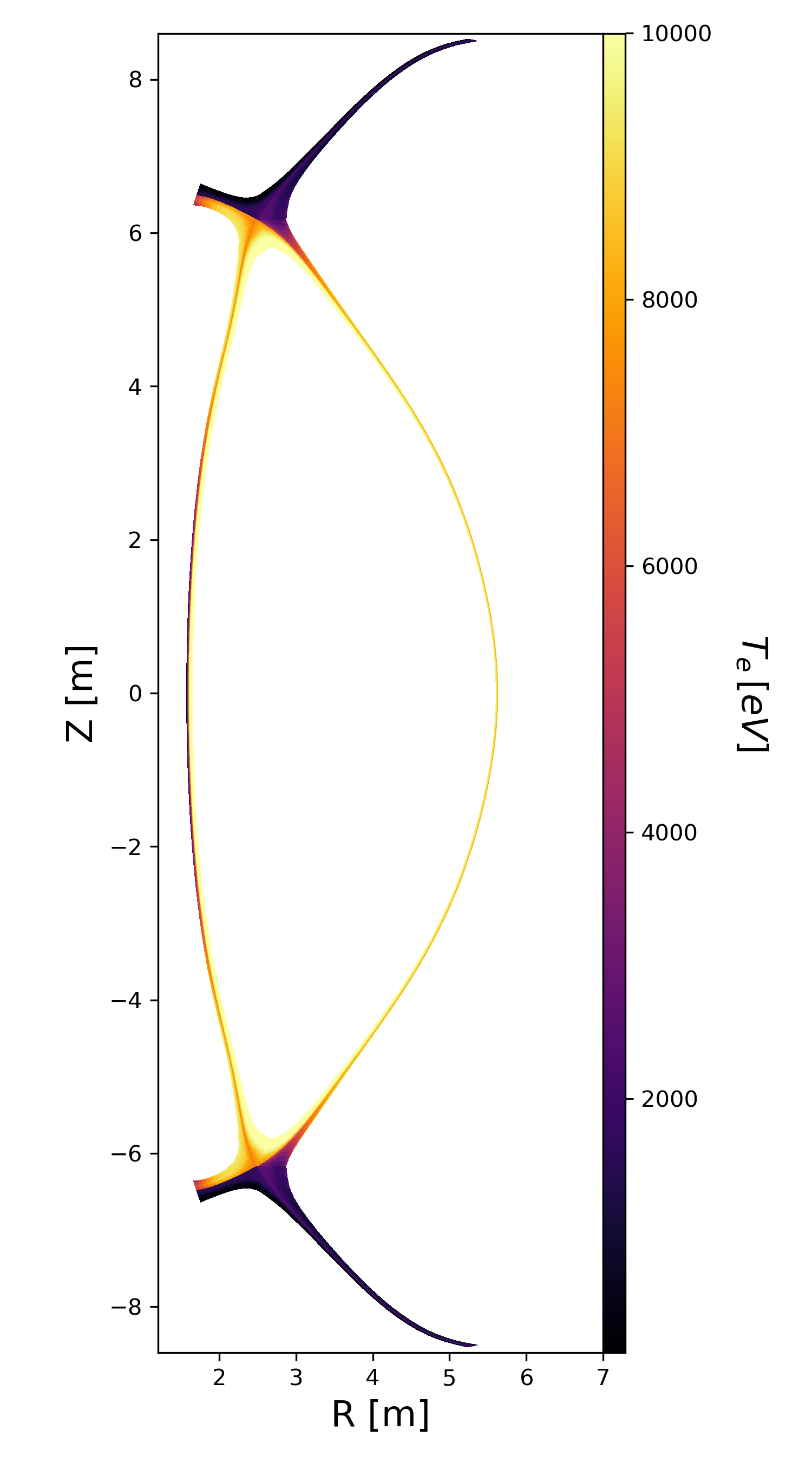}
\includegraphics[width=0.24\textwidth]{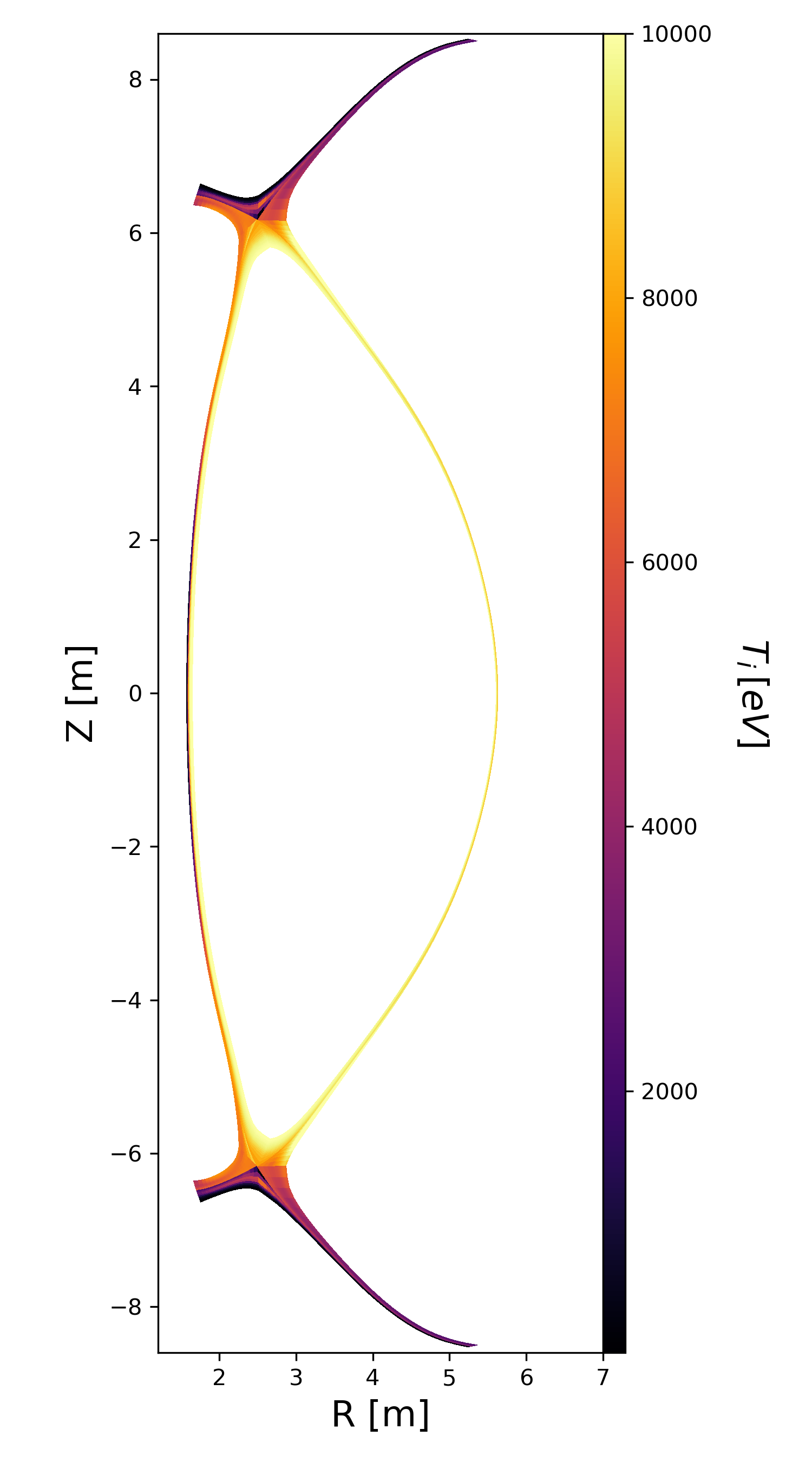}
\caption{Simulation results from the coupled Gkeyll-EIRENE simulation of STEP in a low redycling regime. The electron density, neutral deuterium density, electron temperature, and deuterium ion temperature are plotted in the poloidal plane. The upstream plasma temperature remains at $\sim$9 keV and the downstream density remains low at 1.5$\times$10$^{16}$ $m^{-3}$.}
\label{fig:moments}
\end{figure}

The magnetic geometry features a long outboard divertor leg with tightly packed flux surfaces near the divertor plate. Because the plasma near the divertor plate is narrow, most of neutral deuterium ejected from the plate and walls passes through the plasma without undergoing a reaction. The ionization mean free path of the neutral deuterium is 10s of meters in the  outboard divertor leg, and the SOL-width is about 15 mm, so neutrals pass through the plasma and hit the side walls many times before undergoing a reaction. With a recycling coefficient of 0.5 at the walls, most of the neutrals recycled from the plate are absorbed by the walls before having a chance to react with the plasma. This results in a very low neutral and plasma density near the divertor plates and prevents the neutrals from significantly cooling the plasma. The electron density, neutral deuterium density, electron temperature, and ion temperature are shown in Fig.~\ref{fig:moments}. The upstream density at the OMP is ~$n_{sep}\sim 6\times 10^{17}\, m^{-3}$, and $T_{sep} \sim 9\,keV$.

%\begin{minipage}{0.55\textwidth}
%
%\end{minipage} \hfill
%\begin{minipage}{0.34\textwidth}
%\end{minipage}

\begin{figure}[htbp]
\centering
\includegraphics[width=0.45\textwidth]{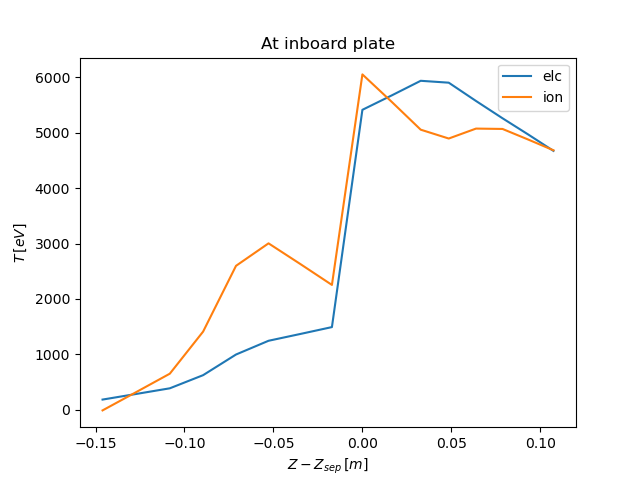}
\includegraphics[width=0.45\textwidth]{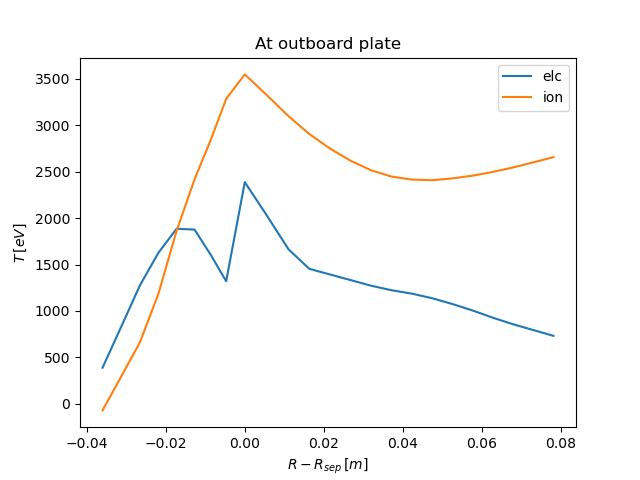}
\includegraphics[width=0.45\textwidth]{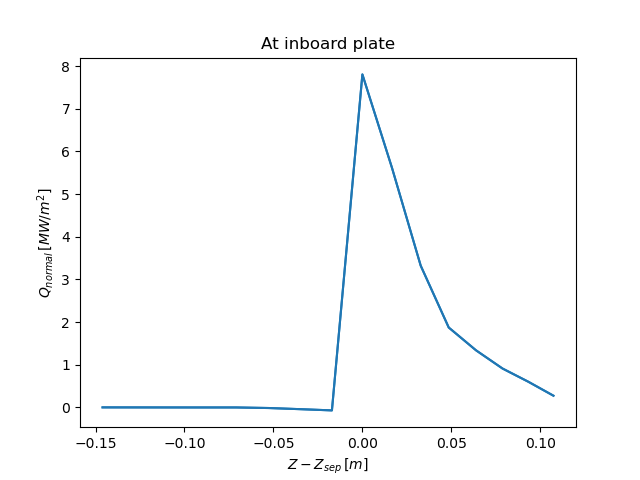}
\includegraphics[width=0.45\textwidth]{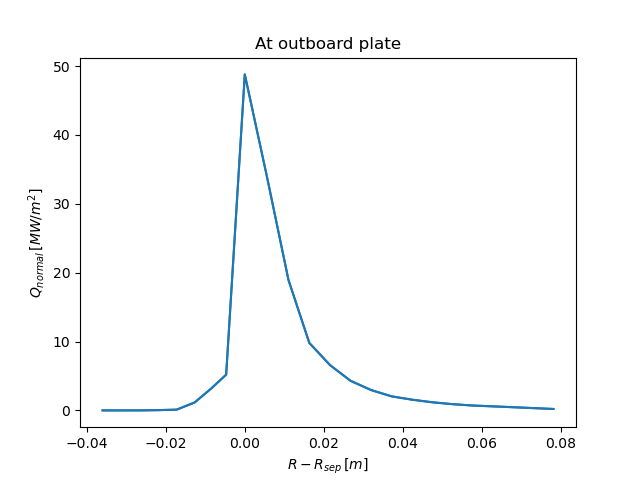}
\caption{
Electron temperature, ion temperature, and total heat flux at the lower inboard and upper outboard divertor plates.
}
\label{fig:plate}
\end{figure}

\vspace{0.5cm}
The total heat flux normal to the plate and the electron and ion temperature at the plate are shown in Fig.~\ref{fig:plate}. The outboard heat flux width mapped upstream is $2.3\,mm$. The peak heat flux on the outboard plate is quite large at 50 $MW/m^2$ and the ions incident at the plate at the peak of the heat flux channel have a temperature of more than 3 keV. 
If the divertor had a lithium coating, these heat fluxes would quickly cause the target to heat up and the lithium to evaporate. 
%Some simple estimates indicate that the evaporated lithium would dominate over the recycled deuterium and cool the SOL, taking it out of the low density, high temperature regime which is advantageous for confinement.
Here we make some simple estimates, which indicate that the evaporated lithium would dominate over the recycled deuterium and cool the SOL. This would take the SOL out of the low density, high temperature regime, thus removing the low recycling conditions that are favorable for confinement.

The total number of deuterium ions incident on one of the outboard plates per second is $1\times 10^{22}$. Using the well known lithium vapor pressure and the plasma-wetted area of the outboard target ($\sim 1.7\,m^2$), one can estimate the number of lithium particles evaporated per second based on the target temperature. For temperatures of $450\, ^\circ C$, $650\, ^\circ C$, and $850\, ^\circ C$, the number of lithium particles evaporating from the target per second would be $5\times 10^{21}$, $1.2\times 10^{24}$, and $3.8\times 10^{25}$ respectively. So, the amount of evaporated lithium will be comparable to or much greater than the recycled deuterium, and the amount of evaporated lithium goes up very quickly with target temperature.
At these low electron densities, only a small portion of the evaporated lithium will be ionized before passing through the plasma and sticking to the wall, but this amount would likely be large enough to significantly cool the SOL and raise the density.
An estimate using the width of the plasma near the divertor plate (15 mm), the electron temperature at the divertor plate ($~\sim 2\,keV$), and the electron density at the divertor plate ($1.5\times 10^{16}\, m^{-3}$) indicates that around 1\% of the lithium will be ionized before passing through the plasma.

So, if the target reaches a temperature of $650\, ^\circ C$, the amount of lithium ionized would be comparable to the total amount of deuterium hitting the plate. In a low recycling regime, the plate temperature will likely be quite high because of the large heat fluxes. For the parameters of this simulation, which has a peak heat flux of 50 $MW/m^2$, one can estimate the approximate target temperature at 730 $^\circ C$ if the plate was tungsten with a 1 mm thick lithium coating.
Thus, a lithium coating on the divertor plate would likely significantly cool the SOL and increase the density. Fortunately, our simulation indicates that even without a low-recycling target material, a hot, low density SOL could be achieved. This would allow a material with a lower vapor pressure to be used on the target.  

However, there are still challenges associated with heat flux handling to address. Even a material like tungsten would be quickly eroded under the conditions in the simulation. A low-recycling solution would likely have to involve a liquid metal target to avoid detrimental erosion. Liquid metals that could potentially be used to coat the divertor target include tin, gallium, aluminum, beryllium, and alloys thereof~\cite{SuperXT}. Additionally, the high ion temperatures will still cause a large degree of sputtering which could contaminate the core. One way to reduce the heat peak heat flux would be to decrease the angle between the field line and target; the current simulation has an angle of 6$^\circ$ but the angle could be realistically reduced to as low as 3$^\circ$ which would reduce the peak heat flux by a factor of 2. Our simulation results presented in section~\ref{sec:drifts} also indicate that mirror trapping can significantly broaden the heat flux width and lower the peak heat flux when drifts are included. In the next section, section~\ref{sec:impurities}, we outline kinetic effects that can potentially address the issue of sputtering and core contamination. 

\section{Confinement of Impurities}
\label{sec:impurities}
\begin{figure}[htbp]
\centering
\includegraphics[width=0.5\textwidth]{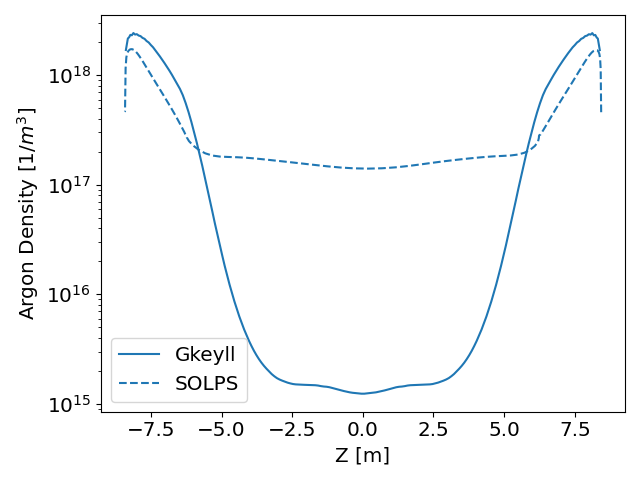}
\caption{Total charged argon density (summed over charge states) plotted along separatrix in the outboard SOL  in SOLPS and Gkeyll. The upstream impurity density in Gkeyll is a factor of 100 lower than in SOLPS~\cite{Shukla25}.
}
\label{fig:impurities}
\end{figure}

In Ref.~\cite{Shukla25} we conducted SOL-only simulations including argon impurities in a more collisional regime than the simulations described in section~\ref{sec:10kev}. The regime was still collisionless enough to observe kinetic effects and differences between gyrokinetic and fluid simulations. The upstream ion temperature was approximately 1 keV in Gkeyll and 0.6 keV in SOLPS; the difference was due to ion mirror trapping in Gkeyll which lowered the heat conduction and raised the ion temperature. We found that, in a magnetic configuration with a Super-X like divertor, the mirror force accelerates particles along the divertor leg resulting in an enhanced potential drop along the field line~\cite{Mike23} and demonstrated that the assumption of equal ion and impurity temperatures often made in fluid codes is not true in collisionless regimes. The combination of the enhanced potential drop and low impurity temperature results in superior confinement of impurities to the divertor region in kinetic simulations; when the potential drop from midplane to divertor plate is large relative to the impurity temperature, the fraction of impurities able to travel upstream is small. We found that this effect, which was only present in the gyrokinetic simulations, drastically reduced the upstream impurity density. Fig.~\ref{fig:impurities} shows the total argon density from a Gkeyll and SOLPS simulation along the outboard SOL at the separatrix. In Gkeyll simulations, electrostatic confinement of impurities to the divertor region led to an upstream argon density 100 times lower than that observed in SOLPS.

The simulations conducted in Ref.~\cite{Shukla25} feature an upstream temperature of around 1 keV and demonstrate the impurity shielding effect. In a true low recycling regime with upstream temperatures of 10 keV or more, the potential drop along the divertor leg will be much larger and the impurity shielding effect would likely be even stronger.
This result addresses another concern about the low-recycling regime. High temperature ions incident at the target will lead to large amounts of sputtering, but if sputtered impurities can be confined to the divertor region as demonstrated in Ref.~\cite{Shukla25}, core contamination can be avoided. 

\section{Heat Flux Broadening: Mirror Trapping \& Drifts}
\label{sec:drifts}
\begin{figure}[htbp]
\centering
\includegraphics[width=0.7\textwidth]{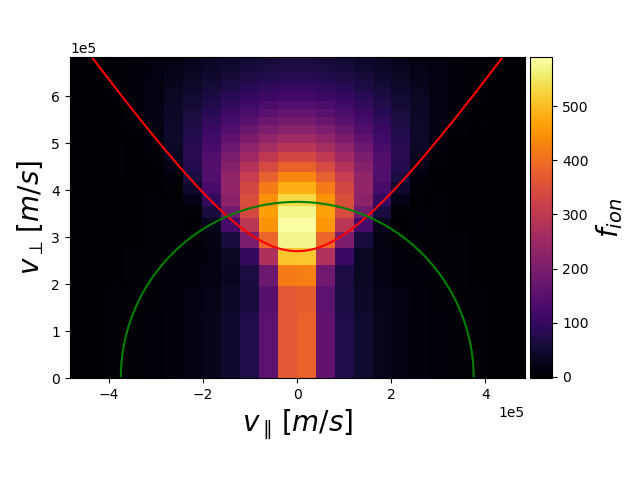}
\caption{ Ion distribution at the OMP. The red line indicates the trapped-passing boundary including the effect of the potential and the green line is a contour of a Maxwellian with the same temperature as the distribution function~\cite{Shukla25}.
}
\label{fig:dist}
\end{figure}

We also conducted a set of SOL-only simulations using both SOLPS and Gkeyll with and without drifts to investigate the effect of drifts on the peak heat flux and heat flux width. These simulations are similar to those discussed in section~\ref{sec:impurities} with an upstream temperature of ~1 keV rather than 10 keV but do not include argon. In these simulations, the magnetic field is halved to emphasize the effect of drifts. We found that while drifts had a negligible effect on the heat flux width in fluid simulations, there was signifcant broadening in kinetic simulations. In kinetic simulations, drifts caused the heat flux channel to spread out to approximately the ion banana orbit width (or poloidal ion gyroradius). This is due to mirror trapping: in a collisionless regime, an ST has a significant trapped fraction on the outboard, and ions spread out radially during banana orbits before being de-trapped and lost to the plate. Because the mirror force is neglected by SOLPS's fluid equations, this effect is only observed in the kinetic simulations. The ion distribution function from Gkeyll at the OMP near the separatrix is shown in Fig.~\ref{fig:dist}. The red line indicates the trapped-passing boundary including the effect of the electrostatic potential, so Fig.~\ref{fig:dist} clearly shows that the ions are trapped upstream.

\begin{figure}[htbp]
\centering
\includegraphics[width=0.45\textwidth]{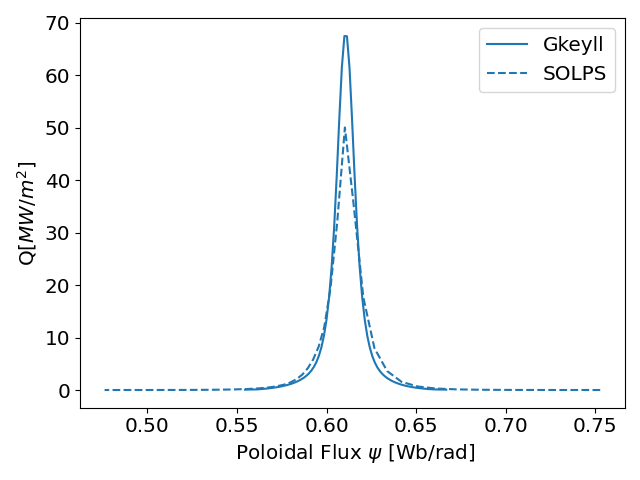}
\includegraphics[width=0.45\textwidth]{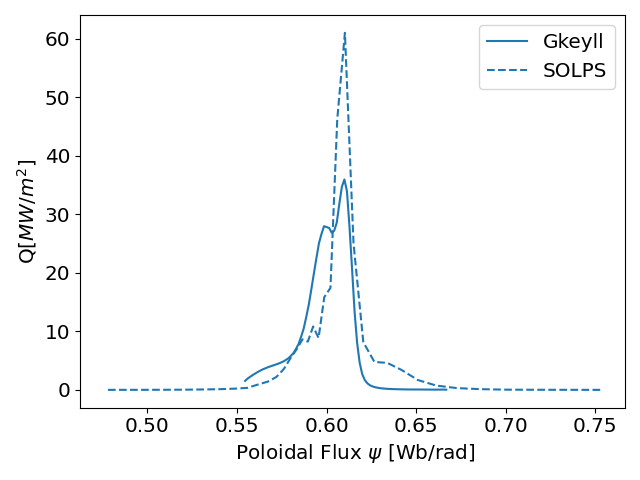}
\caption{Heat Flux at upper outboard plate from SOLPS and Gkeyll simulations without drifts (left) and with drifts (right). Without drifts, both codes have a heat flux width (mapped upstream) of 3mm. With drifts, the SOLPS width remains at 3mm, but the Gkeyll width is increased to the 6.4mm, which is the poloidal ion gyroradius, and the peak heat flux is reduced by a factor of 2.
}
\label{fig:drifts}
\end{figure}

In Fig~\ref{fig:drifts} we plot the heat flux incident at the outboard plate from Gkeyll and SOLPS simulations with and without drifts. The heat flux width in SOLPS is 3 mm with and without drifts; the width is barely affected by the inclusion of drifts. The absence of the experimentally observed dependence of the heat flux width on the poloidal magnetic field has been observed in previous SOLPS simulations and is discussed in Ref.~\cite{Meier2017}. In Gkeyll, the heat flux width is 3 mm without drifts but 6.4 mm (the ion banana width) with drifts and the peak heat flux is reduced by a factor of 2. This result indicates that in a collisionless low-recycling SOL, we can take advantage of the kinetic parallel dynamics to broaden the heat flux channel and reduce peak heat loads. In an even less collisional, hotter SOL, like the 10keV case simulated in section~\ref{sec:10kev}, the ion banana width is 1 cm, so the broadening effect would likely be larger as well.

\section{Conclusion}
\label{sec:conclusion}
Low-recycling regimes present advantages for core confinement but the feasibility of a low-recycling SOL is uncertain. 
%Low recycling materials such as lithium cannot withstand high heat fluxes at the target without evaporating and cooling the plasma, which would interfere with the confinement advantages low-recycling regimes provide.
When using low recycling materials such as lithium as a target material, it is difficult to prevent them from evaporating and cooling the plasma, which would interfere with the confinement advantages low-recycling regimes provide.
Here we show gyrokinetic simulation results that indicate that a low-recycling regime could be achieved by using a low recycling material to coat only the side walls, which receive a much lower heat flux than the target. This is an important step towards demonstrating the feasibility of a low-recycling SOL.

The large target heat fluxes and temperatures present in a low-recycling regime pose challenges such as sputtering and erosion. Here, we note that a low-recycling solution would likely have to involve a liquid metal target to avoid detrimental erosion and also begin to investigate ways to address some of these challenges by taking advantage of kinetic effects. By comparing fluid and gyrokinetic simulations in a moderately collisionless SOL, we find that (1) kinetic effects can help confine impurities downstream, which would help avoid core contamination caused by sputtering and (2) the interaction of mirror trapping and drifts can significantly increase the heat flux width and reduce the peak heat flux.

One important issue we have not yet explored is helium exhaust in a low recycling SOL. If the neutral mean free path is long relative to the pump opening, it may be difficult to pump out helium. However, with a long-legged divertor like the one featured in the STEP magnetic geometry, there could be room to make a very large pump opening to ease the pumping of neutral helium. Additionally, the low electron density in the divertor leg will reduce the helium ionization rate and could make it easier to pump out helium before it reacts with and cools the plasma. In future work, we plan to conduct simulations including helium to address the issue of helium exhaust in a low recycling SOL.

%
% Each of the commands below will create an unnumbered section with the appropriate heading.
% Remove any sections that are not relevant for your article.
% All sections except suppdata will be removed if the [anonymous] option is used.
% See iopjournal-guidelines.pdf for more information.
%

% \ack{Sample text inserted for demonstration.}

\funding{This work has been funded by CEDA SciDAC (Center for Computational Evaluation and Design of Actuators for Core-edge Integration) grant number DE-AC02-09CH11466, Exofusion, and other DOE sources. This research used resources of the National Energy Research Scientific Computing Center, a DOE Office of Science User Facility}
% This section is a list of funder names and grant numbers

% \roles{Sample text inserted for demonstration.}
% % List author names and the contributions made to the article, using terms from the NISO Contributor Roles Taxonomy (CRediT) https://credit.niso.org
% 
% \data{Sample text inserted for demonstration.}
% % For more information on IOP Publishing's research data policy see: https://publishingsupport.iopscience.iop.org/questions/research-data/
% 
% \suppdata{Sample text inserted for demonstration.}

%\section*{References}
\printbibliography

@article{Hudoba2023,
title = {Divertor optimisation and power handling in spherical tokamak reactors},
journal = {Nuclear Materials and Energy},
volume = {35},
pages = {101410},
year = {2023},
issn = {2352-1791},
doi = {https://doi.org/10.1016/j.nme.2023.101410},
url = {https://www.sciencedirect.com/science/article/pii/S2352179123000492},
author = {A. Hudoba and S. Newton and G. Voss and G. Cunningham and S. Henderson},
}

@article{Osawa2023,
doi = {10.1088/1741-4326/acd863},
url = {https://dx.doi.org/10.1088/1741-4326/acd863},
year = {2023},
month = {jun},
publisher = {IOP Publishing},
volume = {63},
number = {7},
pages = {076032},
author = {R.T. Osawa and D. Moulton and S.L. Newton and S.S. Henderson and B. Lipschultz and A. Hudoba},
title = {SOLPS-ITER analysis of a proposed STEP double null geometry: impact of the degree of disconnection on power-sharing},
journal = {Nuclear Fusion},
}

@article{Zhang2024,
author = {Zhang, M. Z. and Sang, C. F. and Zhao, M. L. and Rognlien, T. D. and Zhang, C. and Wang, Y. L. and Bian, Y. and Wang, Y.},
title = {UEDGE modeling of plasma detachment of CFETR with ITER-like divertor geometry by external impurity seeding},
journal = {Contributions to Plasma Physics},
volume = {64},
number = {7-8},
pages = {e202300135},
doi = {https://doi.org/10.1002/ctpp.202300135},
url = {https://onlinelibrary.wiley.com/doi/abs/10.1002/ctpp.202300135},
eprint = {https://onlinelibrary.wiley.com/doi/pdf/10.1002/ctpp.202300135},
year = {2024}
}

@article{Rozhansky2021,
doi = {10.1088/1741-4326/ac3699},
url = {https://dx.doi.org/10.1088/1741-4326/ac3699},
year = {2021},
month = {dec},
publisher = {IOP Publishing},
volume = {61},
number = {12},
pages = {126073},
author = {V. Rozhansky and E. Kaveeva and I. Senichenkov and I. Veselova and S. Voskoboynikov and R.A. Pitts and D. Coster and C. Giroud and S. Wiesen},
title = {Multi-machine SOLPS-ITER comparison of impurity seeded H-mode radiative divertor regimes with metal walls},
journal = {Nuclear Fusion},
}

@article{Shukla25,
    author = {Shukla, A. and Roeltgen, J. and Kotschenreuther, M. and Juno, J. and Bernard, T. N. and Hakim, A. and Hammett, G. W. and Hatch, D. R. and Mahajan, S. M. and Francisquez, M.},
    title = {Direct comparison of gyrokinetic and fluid scrape-off layer simulations},
    journal = {AIP Advances},
    volume = {15},
    number = {7},
    pages = {075121},
    year = {2025},
    month = {07},
    issn = {2158-3226},
    doi = {10.1063/5.0268104},
    url = {https://doi.org/10.1063/5.0268104},
    eprint = {https://pubs.aip.org/aip/adv/article-pdf/doi/10.1063/5.0268104/20586894/075121_1_5.0268104.pdf},
}

@inproceedings{Mike23,
  title = {Physics Based Routes To Increased Confinement},
  author = {Kotschenreuther, M. and Hatch, D.R. and Mahajan, S. and Merlo, G. and Liu, X. and Shukla, A. and Garofalo, A. and Ding, S. and Park, J. M. and Hassan, E. and Wang, Z. and Hu, Y. and Gong, X.},
  year         = 2023,
  month        = {October},
  booktitle    = {29th IAEA 29 Fusion Energy Conference},
  address      = {London, UK},
  organization = {IAEA}
}

@misc{Mana25,
      title={Conservative velocity mappings for discontinuous Galerkin kinetics}, 
      author={Manaure Francisquez and Petr Cagas and Akash Shukla and James Juno and Gregory W. Hammett},
      year={2025},
      eprint={2505.10754},
      archivePrefix={arXiv},
      primaryClass={physics.plasm-ph},
      url={https://arxiv.org/abs/2505.10754}
}

@article{Wiesen25,
title = {The new SOLPS-ITER code package},
journal = {Journal of Nuclear Materials},
volume = {463},
pages = {480-484},
year = {2015},
note = {PLASMA-SURFACE INTERACTIONS 21},
issn = {0022-3115},
doi = {https://doi.org/10.1016/j.jnucmat.2014.10.012},
url = {https://www.sciencedirect.com/science/article/pii/S0022311514006965},
author = {S. Wiesen and D. Reiter and V. Kotov and M. Baelmans and W. Dekeyser and A.S. Kukushkin and S.W. Lisgo and R.A. Pitts and V. Rozhansky and G. Saibene and I. Veselova and S. Voskoboynikov}
}

@article{Schneider2006,
author = {Schneider, R. and Bonnin, X. and Borrass, K. and Coster, D. P. and Kastelewicz, H. and Reiter, D. and Rozhansky, V. A. and Braams, B. J.},
title = {Plasma Edge Physics with B2-Eirene},
journal = {Contributions to Plasma Physics},
volume = {46},
number = {1-2},
pages = {3-191},
doi = {https://doi.org/10.1002/ctpp.200610001},
url = {https://onlinelibrary.wiley.com/doi/abs/10.1002/ctpp.200610001},
eprint = {https://onlinelibrary.wiley.com/doi/pdf/10.1002/ctpp.200610001},
year = {2006}
}

@misc{ARPAE2025,
  title        = {Novel Liquid Metal Plasma Facing Component Alloys},
  author       = {{ExoFusion (ARPA-E)}},
  year         = {2025},
  month        = {April 7},
  howpublished = {ARPA-E project description, “Novel Liquid Metal Plasma Facing Component Alloys”},
  note         = {Developed by ExoFusion under the CHADWICK program; involves designing low-vapor-pressure, low-melting-point liquid metals for continuously replenished fusion first-wall components},
  url          = {https://arpa-e.energy.gov/programs-and-initiatives/search-all-projects/novel-liquid-metal-plasma-facing-component-alloys}
}

@misc{INFUSE2024,
  title        = {Testing Novel Liquid Metal PFC compositions},
  author       = {{ExoFusion (INFUSE)}},
  year         = {2024},
  month        = {August 8},
  howpublished = {INFUSE project description, “Testing Novel Liquid Metal PFC compositions”},
  note         = {Experimental testing of novel liquid metal plasma-facing component alloys at Penn State University’s Radiation Surface Science and Engineering Laboratory},
  url          = {https://infuse.ornl.gov/awards/testing-novel-liquid-metal-pfc-compositions/}
}

@article{Meier2017,
title = {Drifts, currents, and power scrape-off width in SOLPS-ITER modeling of DIII-D},
journal = {Nuclear Materials and Energy},
volume = {12},
pages = {973-977},
year = {2017},
note = {Proceedings of the 22nd International Conference on Plasma Surface Interactions 2016, 22nd PSI},
issn = {2352-1791},
doi = {https://doi.org/10.1016/j.nme.2016.12.016},
url = {https://www.sciencedirect.com/science/article/pii/S2352179116301545},
author = {E.T. Meier and R.J. Goldston and E.G. Kaveeva and M.A. Makowski and S. Mordijck and V.A. Rozhansky and I.Yu. Senichenkov and S.P. Voskoboynikov},
}

@misc{antoine25,
title={Towards fully predictive gyrokinetic full-f simulations}, 
author={A. C. D. Hoffmann and T. N. Bernard and M. Francisquez and G. W. Hammett and A. Hakim and J. Boedo and R. Rizkallah and C. K. Tsui and the TCV team},
year={2025},
eprint={2510.11874},
archivePrefix={arXiv},
primaryClass={physics.plasm-ph},
url={https://arxiv.org/abs/2510.11874}, 
}

@misc{Shukla25Xpt,
      title={Constructing Field Aligned Coordinate Systems for Gyrokinetic Simulations of Tokamaks in X-point Geometries}, 
      author={Akash Shukla and Ammar Hakim and James Juno and Gregory Hammett and Manaure Francisquez},
      year={2025},
      eprint={2510.21676},
      archivePrefix={arXiv},
      primaryClass={physics.plasm-ph},
      url={https://arxiv.org/abs/2510.21676}, 
}

@patent{SuperXT,
  author       = {Michael Kotschenreuther},
  title        = {Increasing Energy Gain in Magnetically Confined Fusion Plasmas by Increasing the Edge Temperature: The Super-XT Divertor},
  assignee     = {Michael Kotschenreuther},
  number       = {US20230245792A1},
  year         = {2023},
  month        = {8}, 
  day          = {3},
}

\end{document}